\documentclass[twocolumn,showpacs,amsmath,amssymb]{revtex4}

\usepackage{graphicx}
\usepackage{dcolumn}
\usepackage{bm}

\newcommand{\be}{\begin{equation}}
\newcommand{\ee}{\end{equation}}
\newcommand{\bfig}{\begin{figure}}
\newcommand{\efig}{\end{figure}}
\newcommand{\incl}{\includegraphics}

\begin{document}
\title{The anomalous Hall Effect and magnetoresistance in the layered ferromagnet Fe$_{\frac14}$TaS$_{2}$: the inelastic regime.}
\author{J. G. Checkelsky$^1$, Minhyea Lee$^1$, E. Morosan$^2$, R. J. Cava$^2$  and N. P. Ong$^1$}
\affiliation{$^1$Department of Physics and $^2$Department of Chemistry, Princeton University, Princeton, NJ 08544, USA}

\date{\today}
\begin{abstract}
The large magnetic anisotropy in the layered ferromagnet Fe$_{1/4}$TaS$_2$ leads
to very sharp reversals of the magnetization $\bf M$ at the coercive field.  
We have exploited this feature to measure the anomalous Hall effect (AHE), focussing on
the AHE conductivity $\sigma^A_{xy}$ in the inelastic regime.  At 
low temperature $T$ (5-50 K), $\sigma^A_{xy}$ is $T$-independent, consistent with the Berry-phase/Karplus-Luttinger theory.
Above 50 K, we extract an inelastic AHE conductivity $\sigma^{in}_{xy}$
that scales as the square of $\Delta\rho$ (the $T$ dependent
part of the resistivity $\rho$).  The term $\sigma^{in}_{xy}$ clarifies the 
$T$ dependence and sign-reversal of the AHE coefficient $R_s(T)$.  
We discuss the possible ubiquity of $\sigma^{in}_{xy}$ in 
ferromagnets, and ideas for interpreting its scaling with $(\Delta\rho)^2$.  
Measurements of the magnetoresistance (MR) reveal a rich pattern of behavior vs. $T$ and field tilt-angle.  We show that the 2 mechanisms, the anisotropic MR effect and field-suppression of magnons, account for the intricate MR behavior,
including the bow-tie features caused by the sharp reversals in $\bf M$.
\end{abstract}

\maketitle                   

\section{Introduction}\label{intro}
In a ferromagnet, the appearance of spontaneous magnetization breaks 
time-reversal symmetry (TRS) in the spin degrees of freedom.  
Spin-orbit coupling conveys the loss of TRS to the charge degrees.
Hence the appearance of spontaneous magnetization strongly influences the electrical 
currents.  Although the investigation of this topic has had a very long
history, interest in its many ramifications continues to 
surface, as our understanding of quantum effects
in electron transport improves. The past 7 years have seen strong resurgent interest
-- both theoretical~\cite{Ye,Niu,Nagaosa,Tatara,Jungwirth,Yao,Haldane,Onoda} 
and experimental~\cite{Matl,Tokura,WLee,Mathieu,Manyala,Gil,Zeng06,Onose,mlee07} --
on the anomalous Hall effect (AHE), which is perhaps 
the most fascinating manifestation of TRS breaking in a ferromagnet.  
Recent research has clarified the fundamental relation between
the Berry phase and the anomalous velocity which leads directly to the 
AHE when time-reversal invariance is broken.  The notion of an
anomalous velocity in a lattice has deep 
roots in the physics of solids, starting with the seminal 1954
theory of Karplus and Luttinger (KL)~\cite{Karplus,Blount}.  
The modern interest is also fueled by experiments to reverse
magnetization by current in spin-based devices.  These experiments 
explore anew issues pertaining to domain wall motion in applied currents,
and the reciprocal effects of domain motion on transport~\cite{Maekawa}.

We report transport experiments on the layered
dichalcogenide ferromagnet Fe$_{\frac14}$TaS$_2$, in which the 
spontaneous magnetization is strongly pinned perpendicular to 
the TaS$_2$ layers by a very large anisotropy field at temperatures $T$ below 
the Curie point $T_C$ (160 K)~\cite{Morosan}.  
In the hysteresis loops measured up to 100 K, 
reversal of $\bf M$ at the coercive field occurs as a very sharp jump $\Delta M$, 
which induces a large jump $\Delta\sigma_{xy}$
in the Hall conductivity~\cite{Morosan}.  The ratio of the 
2 quantities enable the anomalous Hall conductivity (AHC) to be accurately 
extracted.  This allows us to address a major problem 
in the AHE in ferromagnets -- the role of inelastic scattering at
elevated $T$.  Our analysis finds that the AHC is the sum of 2 terms of opposite signs.
The first term, an intrinsic term independent of carrier lifetime, is identified
with the Berry-phase/KL term.  The second, dominant at high temperatures 
but negligible below 50 K, arises from scattering by inelastic excitations
-- magnons and thermally excited textures in the magnetic order parameter. 
We isolate the inelastic term and show that it scales as the square
of the inelastic part of the resistivity.  
The complicated, non-monotonic $T$ dependence of the anomalous Hall coefficient $R_s(T)$ 
is seen to be a simple consequence of competition between the Berry phase/KL
term and the inelastic term.  
This competition seems to underlie the temperature profile of
the AHE coefficient in many ferromagnets.

In ferromagnets, the magnetoresistance (MR) is dominated by 2 mechanisms~\cite{McGuire,Gil}.  
One is the anisotropic magnetoresistance (AMR) effect in which 
the scattering rate for electrons with velocities $\bf v\parallel M$ is higher than for $\bf v\perp M$~\cite{Smit,Fert,McGuire}.  The second is the field-suppression of magnons.  
The strong pinning of $\bf M$ to the $c$-axis in Fe$_{\frac14}$TaS$_2$
leads to a rich assortment of MR behavior apparent in field-tilt experiments.  
We show that the 2 mechanisms account very well for the full MR behavior, including
the appearance of ``bow-tie'' features caused by the abrupt reversals in $\bf M$.
The analysis is considerably simplified because the 2 mechanisms dominate in opposite regimes of tilt angles and $T$.
Both imply that scattering from spin excitations are dominant below $T_C$.

\section {Experimental Details}\label{expt}
In the dichalcogenide TaS$_{2}$, the weak van der Waals forces between 
adjacent TaS$_{2}$ layers allows intercalation of most of 
the 3$d$ transition-metal elements~\cite{Parkin}. 
In the system Fe$_x$TaS$_2$, the ground state evolves from 
superconductivity to ferromagnetism with increasing $x$.  
A small Fe content ($x=0.05$) leads to a slight rise in the
superconducting transition temperature $T_c$, but superconductivity
is eventually destroyed~\cite{Dai}.
At large $x$, the Fe ions order in a superlattice.  
The Curie transition to the ferromagnetic state occurs at $T_C$ = 40--160 K.
In the specific interval $x$ = $\frac14 \rightarrow\frac13$, the easy
axis of magnetization is perpendicular to the TaS$_{2}$ layers
~\cite {Parkin}.  We focus on the composition $x = \frac14$ which 
has a very large magnetic anisotropy.

Single crystals of  Fe$_{\frac14}$TaS$_2$ were
grown by iodine-vapor transport (see Ref.~\cite {Morosan} 
for details on growth and sample characterization). 
The magnetization $M$ was measured in a 
SQUID (superconducting quantum interference device) magnetometer. 
In the trasport experiments, several crystals of typical
size $\sim$  1 $\times$ 0.2 mm$^2$ and thickness $\leq$ 20
$\mu$m  were investigated using the standard 4-probe ac lock-in
technique. Electrical contacts, made by silver paint, had typical contact
resistances smaller than 1 $\Omega$.  Rotation of the samples in a field
were performed by a home-built rotation stage, which was in direct 
contact with the cold finger in the cryostat.  
The stage was suspended by sapphire $V$-jewels
strung by Kevlar lines to minimize heating during rotation. 
High-field measurements to
33 T were performed at the National High Magnetic Field Lab., Tallahassee.

\section{Resistivity and Magnetization}
\begin{figure}
\incl[width=9.5cm]{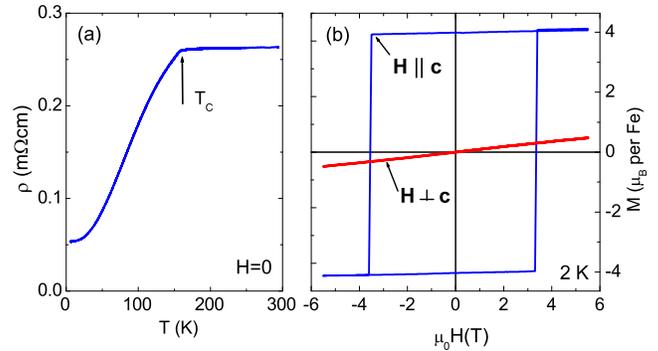}
\caption{\label{MR} (Color online)
(a) The in-plane resistivity $\rho$ vs. $T$ in Fe$_{1/4}$TaS$_2$ 
($H=0$). At the Curie temperature $T_C$ (arrow), $d\rho/dT$ 
has a discontinuity. 
(b) Curves of the magnetization $M$ vs. $H$ at 2 K with
$\bf H\parallel c$ and $\bf H\perp c$.  The former shows a square
shape with vertical jumps at the coercive fields $H_c$ while the latter
is small and $H$-linear.
}
\end{figure}

We first discuss the $T$ dependence of the in-plane 
resistivity $\rho$ and magnetization $M$.  
Figure \ref{MR}a shows that, above the Curie
temperature $T_C$ = 160 K, $\rho$ is nearly 
$T$ independent.  Below $T_C$, however, it
drops by a factor of 4 from 160 K to 4 K.  

In the magnetic dichalcogenides Cr$_{\frac13}$NbS$_2$ and 
Fe$_{\frac14}$NbSe$_2$ ~\cite{mlee_unpulished},
$\rho$ is observed to vary smoothly across $T_C$.  By contrast,
the derivative $d\rho/dT$ in Fe$_{\frac14}$TaS$_2$ displays a sharp discontinuity at $T_C$, which is likely
a consequence of the unusually large magnetic anisotropy.  Below $T_C$,
$\bf M$ is strong pinned to the easy axis $\bf c$ (normal to the 
layers).  As shown in Fig.\ref{MR}b, the hysteretic $M$-$H$ loop 
measured at 2 K has a rectangular shape with near-vertical jumps in $M$ 
occuring at the coercive field $H_c$.  In Panel (b), the straight
line with small slope represents the in-plane magnetization $M_{ab}$ 
induced by $\bf H\perp c$.  The linear increase in $M_{ab}$ implies that the in-plane susceptibility $\chi_{ab}$ is $H$ independent up to 6 T.
This $H$-independent $\chi_{ab}$ will prove useful in the analysis
of the field-tilt MR.  Further, assuming that the linearity in $M_{ab}$ persists to 
intense $H$, we estimate that the anisotropy field $H_{A}\sim$ 60 T at 2 K.  
The jumps in $M$ are observed to $T\sim$ 100 K \cite{Morosan}.


\bfig 
\incl[width=9.5cm]{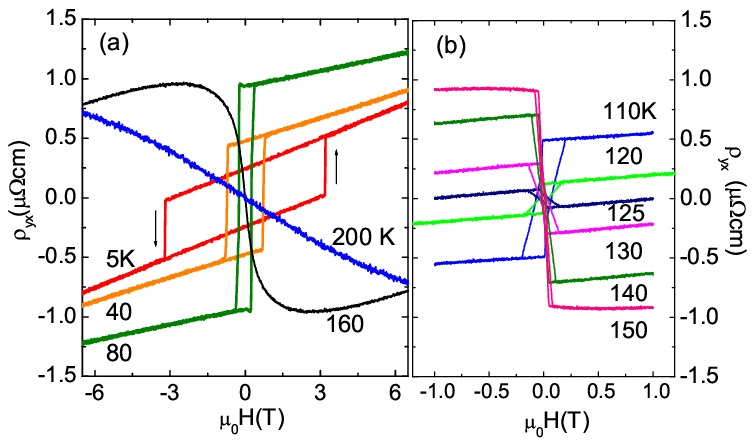} 
\caption{\label{rhoxy} 
(a) The Hall resistivity $\rho_{yx}$ vs. $H$ measured with $\bf H\parallel c$
and $\bf I\perp c$ at $T$ = 5, 40, 80, 160 and 200 K.  Jumps in $\rho_{yx}$ 
occur at $H_c$ in response to the abrupt sign-reversal of $\bf M$. 
Above $T_C$ ($T$ = 160 K and 200 K), $\rho_{yx}$ is still dominated
by the AHE term. 
(b) Expanded vies of hysteresis loops of $\rho_{yx}$ vs. $H$ for $T$ near $T_C$.  Note the reversal in circulation sense between 120 and 125 K. } 
\efig

\section{The Anomalous Hall effect}\label{secAHE}
In the ferromagnetic state of Fe$_{\frac14}$TaS$_2$, the Hall effect 
is a superposition of a term associated with the sharp jumps in $M$ and an
$H$-linear term associated with the Lorentz force. 
Figures \ref{rhoxy}a and \ref{rhoxy}b
display the Hall resistivity ($\rho_{yx}$) in a field $\bf H\parallel c$
with the current $\bf I$ in the $ab$ plane.
Clearly, the hysteresis loop of $\rho_{xy}$ vs. $H$
at 5 K reflects the square magnetic hysteresis
loop shown in Fig. \ref{MR}b.  At $H = H_c$, $\rho_{yx}$ suffers an
abrupt sign-reversal similar to that in $M$, but is otherwise linear
in $H$.  By long practice, $\rho_{yx}$ in a ferromagnet is 
empirically expressed as~\cite{Smith,Hurd}
\be
\rho_{yx} = R_HB + \mu_0 R_sM, 
\label{rxy}
\ee
where $\mu_0$ is the vacuum permeability and
$R_H$ and $R_s$ are the ordinary and anomalous Hall
coefficients, respectively.  The first term is the ordinary 
Hall effect (OHE), while the second term $R_sM$ is referred to as
the anomalous Hall resistivity ($\rho_{yx}^A$).

\bfig
\incl[width=8cm] {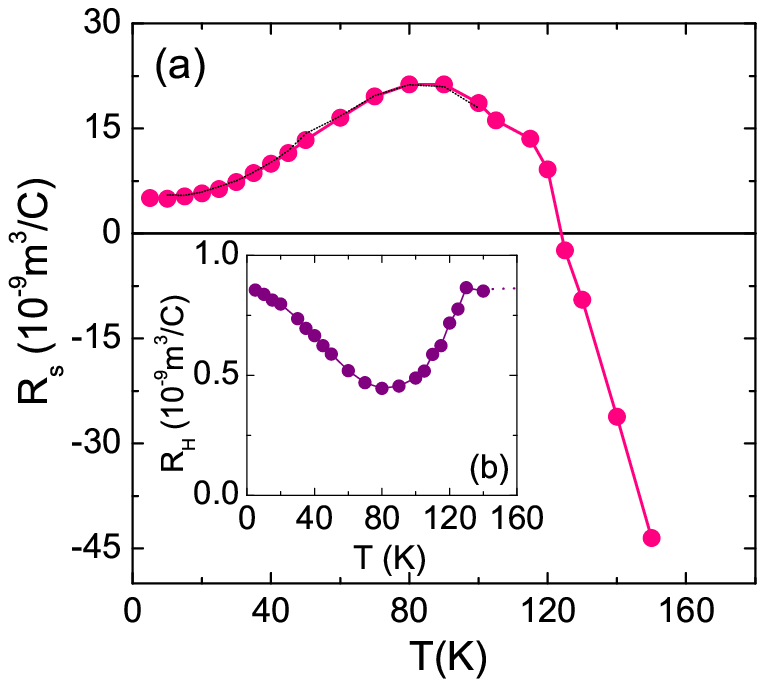}
\caption{\label{RsRH} (Color online)
 (a) The $T$ dependence of the AHE coefficient $R_s$ inferred from
 the jump in $\rho_{xy}$ at $H_c$.
(b) The OHE coefficient $R_H$ vs. $T$ inferred from the $H$-linear
portions of $\rho_{xy}$. 
}
\efig

In the elemental ferromagnets Fe and Ni, the AHE term is so much
larger than the OHE term that one rarely worries about the latter~\cite{Hurd}.
However, in many magnetic systems of current interest (or in pure
samples with very long electron mean-free-path $\ell$), the OHE term
is not negligible, and often comparable in size.  Then the accurate
separation of the 2 terms in Eq. \ref{rxy} poses a difficult experimental problem.  
In Ref. \cite{mlee07}, a method, based on scaling the 
MR curve against the $M$-$H$ curve, was introduced for MnSi which has a very
long $\ell$ at low $T$.  Here, the abruptness of the jump in $M$
provides yet another way to execute this separation.  As we tune $H$ across
$H_c$, inducing an abrupt jump and sign-change in $M$, we engender a 
corresponding jump and sign-change in $\rho_{xy}$.  The 
ratio of the jump magnitudes $\Delta M$ and $\delta\rho_{xy}$ 
is a direct measurement of the AHE coefficient $R_s$, 
with minimal experimental uncertainty.
The squareness of the $M$-$H$ loop provides the most direct
way to determine $R_s$ in ferromagnets with a sizeable OHE term.
Away from the jump, the linear variation of $\rho_{xy}$ with $H$
is used to determine $R_H$.  With this approach, we may reliably
determine $R_s$ and $R_H$ at each $T$ below $T_C$.

As shown in Fig. \ref{RsRH}a, the OHE coeffieint $R_H$
is hole-like, and shows only a moderate $T$ dependence
with a broad minimum at $\sim$80 K.  The Hall number density $n_H = 
1/eR_H$ varies from $8 \times 10^{21}$ cm$^{-3}$ at 5 K and to
the peak value $1.4 \times 10^{22}$ cm$^{-3}$ at 80 K.

How the anomalous Hall resistivity $\rho_{xy}^A$ changes with $T$ may be readily
``read off'' the loops by identifying the jump 
amplitude $\delta\rho_{xy}$ with $2\rho_{xy}^A$ (Figs. \ref{rhoxy}a and b).
At 5 K, $R_s$ starts out positive -- it circles the hysteresis loop 
in the same anticlockwise sense as $M$ (arrows in Fig. \ref{rhoxy}a).  As
$T$ is raised, $\Delta\rho_{yx}$ increases rapidly.  However,
the circulation changes sign near 125 K, above which $\rho_{xy}^A$ 
becomes large and negative.  The inferred anomalous 
Hall coefficient $R_s = \rho_{xy}^A/\mu_0M$ 
is plotted in Fig. \ref{RsRH}b.  Between 5 and 80 K, $R_s$ increases by
a factor of 6 to a broad maximum at 80 K. Then it plunges to large
negative values as $T\rightarrow T_C^{-}$, changing sign at $\sim$125 K.
The peaking of $R_s$ and the sign-reversal are common features 
of the $R_s$-$T$ profile in many ferromagnets~\cite{Hurd,Kats}.  
However, there has been scant progress in understanding its
causes.  We address this point in Sec. \ref{inelastic}.

Above $T_C$, $\rho_{yx}$ is free of hysteresis.  However, the
negative sign of $d\rho_{yx}/dH$ for $T>T_c$ implies that
$\rho_{yx}$ is still dominated by the AHE term $R_s$.  Despite the
vanishing of the spontaneous $M$, field-induced alignment of the moments 
produces a large anomalous Hall response in the paramagnetic state (this is commonly
observed in ferromagnets, e.g. in manganites~\cite{Matl}).

\bfig 
\incl[width=8.5cm]{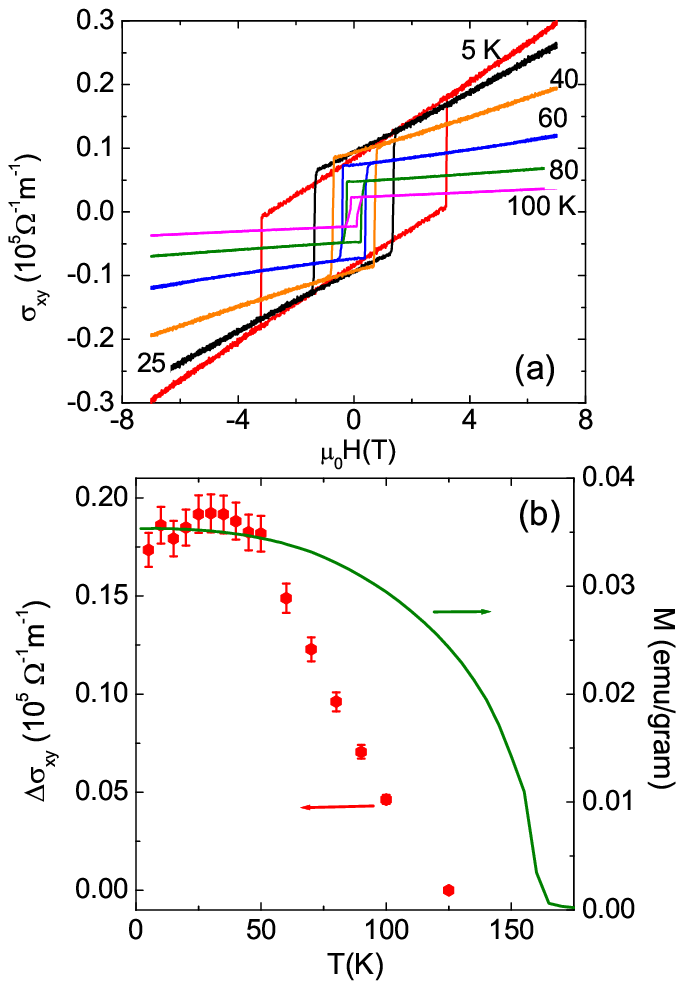}
 \caption{\label{figsxy} (Color online)
(a) Hysteresis loops of $\sigma_{xy}$ (calculated from curves of $\rho_{yx}(H)$ and $\rho(H)$) at selected $T$.  The $H$-linear
segments are the classical Lorentz component while the jumps
are the AHC $\sigma^A_{xy}$.   
(b) The $T$ dependence of the jump magnitude $\Delta\sigma_{xy} = 2 \sigma^A_{xy}$ (solid circles).  For comparison, we also
plot $M$ measured at 0.1 T (solid curve). Note that the
jump magnitude is $T$ independent below 50 K within our resolution. 
Above 50 K, it falls steeply to negative values.
} 
\efig

\subsection{Intrinsic AHE Conductivity at low T}\label{intAHE}
In recent experimental approaches to the AHE problem, one prefers to 
focus on the Hall conductivities which have the  
useful property of additivity (unlike $\rho_{xy}$).  
This view is emphasized in, for e.g., Ref.~\cite{Manyala,Zeng06,Onose,mlee07}. 
The total Hall conductivity $\sigma_{xy}$ is the sum of 
the ordinary Lorentz-force Hall conductivity $\sigma^n_{xy}$
and the anomalous Hall conductivity (AHC) $\sigma^A_{xy}$, 
\be
\sigma_{xy} = \sigma^n_{xy} + \sigma^A_{xy}.
\label{sxy}
\ee

As the AHC scales with the magnetization $M$, we express
it as~\cite{mlee07} 
\be
\sigma^A_{xy} = S_H M, 
\label{sigmaA}
\ee
where the scaling coefficient $S_H$ has
dimensions of Volt$^{-1}$. In MnSi, $S_H$ is shown to be a constant
below $T_C$~\cite{mlee07}.  Multiplying Eq. \ref{sxy} across by $\rho^2$, 
and identifying $\sigma^n_{xy}$ with $R_H\rho^2 H$, we have
\be
\rho_{yx} = R_0 B + \mu_0 S_H\rho^2M, 
\label{rhoH}
\ee
which is Eq. \ref{rxy} with 
\be
R_s = S_H\rho^2.
\label{Rs}
\ee  
The conductivity-additivity viewpoint emphasizes
the constancy of the scaling parameter $S_H$, in contrast to 
$R_s$, which conflates the strong $T$ and $H$ dependences of the resistivity
and $\sigma^A_{xy}$.  This change of perspective involves more
measurements, but it makes comparisons with quantities calculated in
linear-response theory more direct (to theorists, $R_s$ is a complicated
empirical parameter that they usually ignore).

Adopting this approach, we display in Fig. \ref {figsxy}a the
hysteresis loops of the total Hall conductivity $\sigma_{xy}$ 
inferred from the curves of $\rho_{yx}$ and $\rho$ (Fig.\ref{MR}). 
As mentioned, the sharp jumps $\Delta M$ provide direct measurements
of two Hall conductivities in Eq. \ref{sxy}.  At each $T$, we identify
the $H$-linear segments with $\sigma^n_{xy}$, and the jump magnitudes 
with $2 \sigma^A_{xy}$.

Figure \ref{figsxy}b plots the $T$ dependence of 
$\sigma^A_{xy}$ from 5 to 125 K as solid circles.  For comparison,
we have also plotted the $M$ vs. $T$ curve measured in a field of 0.1 T.  
There are 2 noteworthy features.  Below 50 K, both the AHC and $M$ are
only very weakly $T$ dependent, so they may be scaled together
to give an estimate of $S_H$ = $(1.93\pm0.07)
\times 10^4$ V$^{-1}$ between $M$ and $\Delta\sigma_{xy}$.  This
value is about 3.6 times smaller than in MnSi 
(where $S_H\sim 7.04 \times 10^4$ V$^{-1}$)~\cite{mlee07}.
The constancy of the AHC below 50 K is consistent
with the Karplus-Luttinger (KL) prediction.  The existence of the
Berry-phase/KL AHC has now been established in several experiments
~\cite{WLee,Mathieu,Zeng06,Onose,mlee07}.  For a simplified explanation
of the Berry-phase/KL term, see Ref. \cite{OngLee}.

\subsection{AHE in inelastic regime}\label{inelastic}
The second important feature in Fig. \ref{figsxy}b is
the sharp downwards deviation of $\sigma^A_{xy}$
from $M$ above 50 K. This deviation contrasts with
the case in MnSi in which $\sigma^A_{xy}$ is observed
to track the curve of $M$ right up to $T_C$
(29 K)~\cite{mlee07}.  The relatively sharp onset of the
deviation here implies that a distinct 
contribution, negative in sign, appears at 50 K
and grows rapidly in magnitude.  This is rendered quite
apparent if we plot the $T$ dependence of the ratio $\sigma^A_{xy}/M$ (to remove
the $T$ dependence of $M(T)$) (Fig. \ref{sAvsT}a).  
The constancy of the ratio from 5 to 50 K 
(discussed in the previous section) gives way to a steep
decrease above 50 K, consistent with the appearance of a negative contribution
to the AHC (shaded region).

\bfig
\incl[width=9cm] {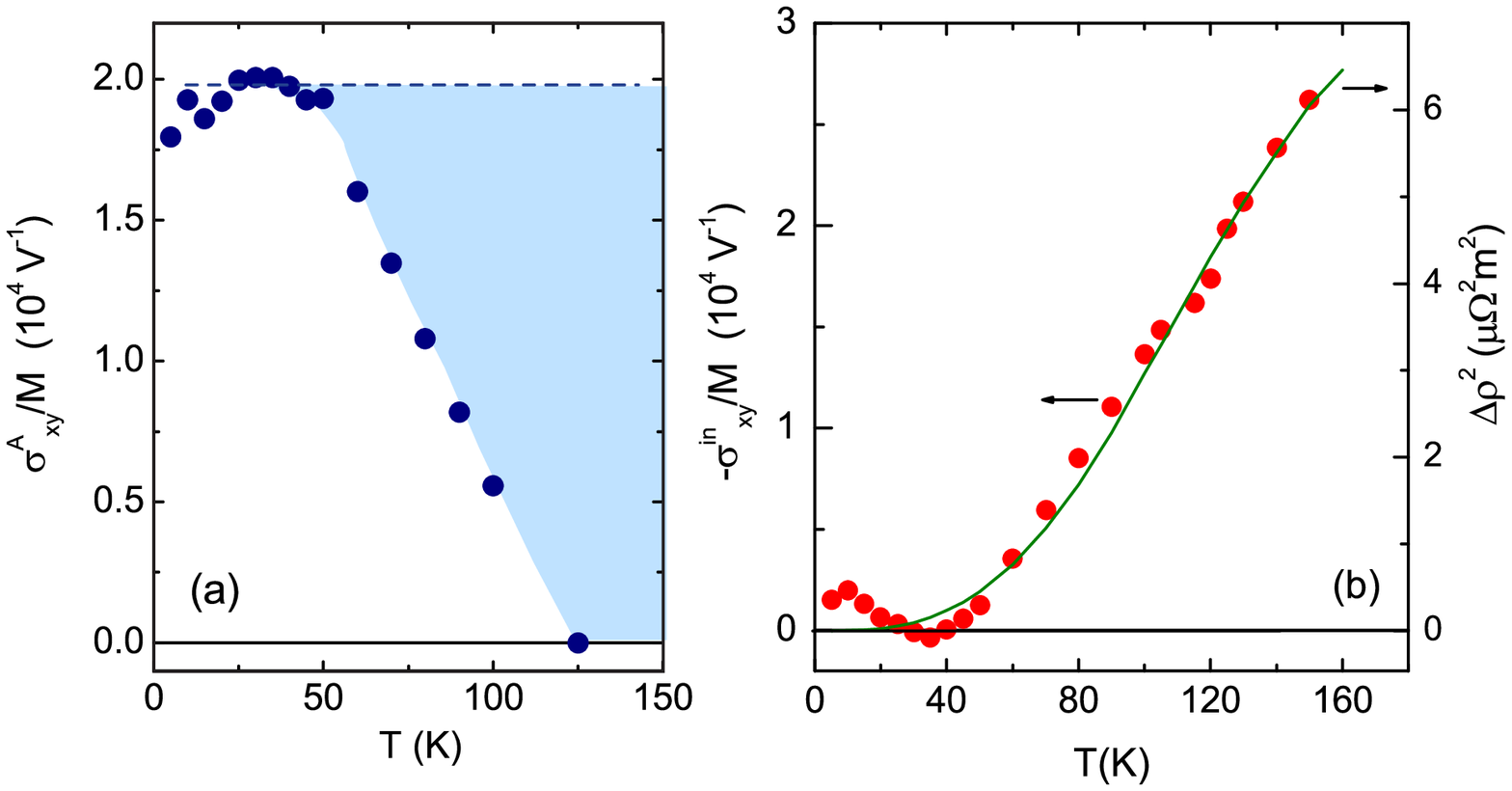}
\caption{\label{sAvsT} (Color online)
The $T$ dependence of the anomalous Hall conductivity
divided by magnetization $\sigma^A_{xy}/M$ (Panel a)
and comparison of the inelastic component 
$\sigma^{in}_{xy}/M$ with $(\Delta\rho)^2$ (Panel b).
In Panel (a), $\sigma^A_{xy}/M$ is expected to be the constant $S_H$
(dashed line).  However, above 50 K, a negative contribution $\sigma^{in}_{xy}/M$ appears and increases rapidly 
in magnitude (shaded region).  $\sigma^{in}_{xy}$ is the 
Hall conductivity produced by inelastic excitations.
Panel (b) shows that the $T$ dependence 
of $\sigma^{in}_{xy}/M$ (solid circles)
matches the square of $\Delta\rho$ (inelastic part of 
resistivity, solid curve).
}
\efig

As the new term's magnitude increases monotonically with $T$, we identify
it with a Hall conductivity $\sigma^{in}_{xy}$ caused by scattering from 
inelastic excitations, notably magnons and spin defects or textures
in the uniform magnetization.  To include this term, the AHC in Eq. \ref{sigmaA} 
(divided by $M$) becomes
\be
\frac{\sigma^A_{xy}}{M} = S_H + \frac{\sigma^{in}_{xy}}{M},
\label{sAM}
\ee
where the 2 terms on the right have opposite signs.  Subtracting
off the constant $S_H$, we display the $T$ dependence of the new term
$-\sigma^{in}_{xy}/M$ in Fig. \ref{sAvsT}b.  This brings out
the monotonic increase in $-\sigma^{in}_{xy}/M$ which extends from 50 K
to $\sim T_C$. In magnitude, $|\sigma^{in}_{xy}|/M$ is equal to $S_H$ near 125 K,
but continues to grow to $\sim$1.5 $S_H$ at $T_C$.  
To underscore its origin in inelastic excitations, we 
have compared it with the inelastic part of the resistivity
$\Delta\rho(T) = \rho(T)-\rho(0)$ (the MR is negligible in
the geometry with $\bf H\parallel c$).  Remarkably,
$-\sigma^{in}_{xy}/M$ matches very well the curve of $(\Delta\rho)^2$
(solid curve in Panel b).  Scaling to $(\Delta\rho)^3$ is much less satisfactory.

We remark that the isolation of the term $\sigma^{in}_{xy}$ (Fig. \ref{sAvsT})
rests on the sole assumption that the Berry-phase/KL term $S_H$ is $T$-independent
up to $T_C$, which has experimental support from Refs.~\cite{Zeng06,Onose,mlee07}.
Independent of this assumption, the existence of a large inelastic term in the AHC 
is immediately apparent from inspection of the raw data 
of $\rho_{xy}$ (Fig. \ref{rhoxy}).  
The jumps $\delta\rho_{xy} = 2\sigma^A_{xy}/\rho^2$ are seen to remain
at a large value even though $M$ decreases rapidly 
as $T\rightarrow T_C^{-}$ (see curves at 130, 140 and 150 K). This requires
an AHC term that is large and negative.
We discuss this further in Sec. \ref{discuss}.

\bfig 
\incl[width=7cm] {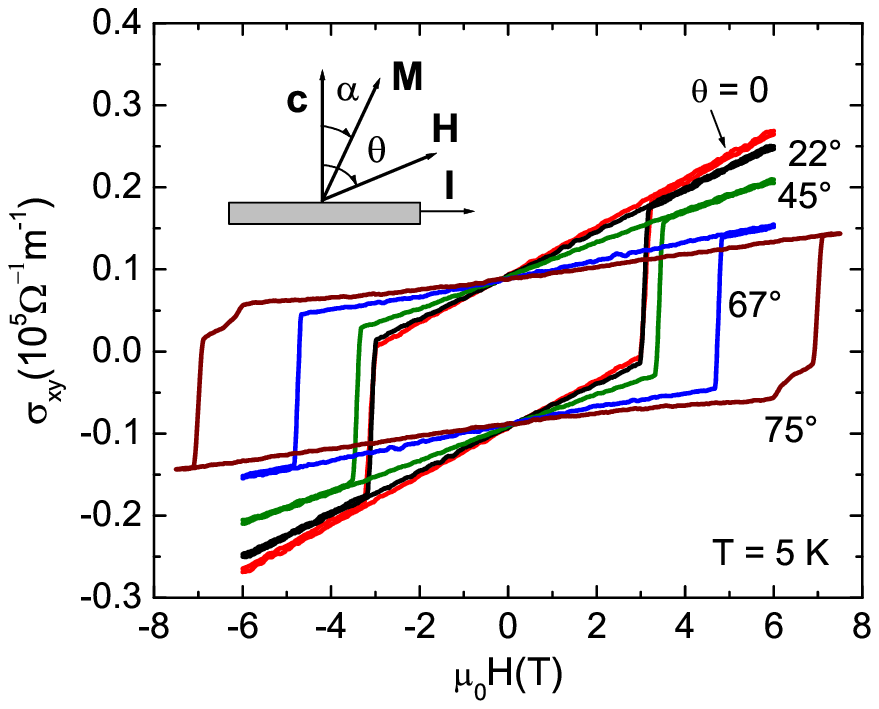}
\caption{\label{sxyangle} (Color online)
Hysteresis loops of $\sigma_{xy}$ at $T$ = 5 K at 
selected field tilt-angles $\theta$.  Although the 
$H$-linear segments (OHE component) changes as $\cos\theta$ (scaling
as the $\bf H\cdot c$), the jump magnitude (AHE component)
is nominally independent of $\theta$.
The inset defines the tilt angles $\alpha$ and $\theta$ relative to $\bf c$
of $\bf M$ and $\bf H$, respectively.
} 
\efig

As mentioned in Sec. \ref{intro}, a difficult aspect of the
AHE problem is the issue inelastic excitations.  To complicate matters,
the $T$ dependence of the AHE, by long convention, is usually reported
in terms of $R_s(T)$, which mixes the $T$ dependences of $\rho$ 
and $\sigma^A_{xy}$ (Sec. \ref{intAHE}). In a broad class of ferromagnets, the 
profile of $R_s(T)$ follows a common pattern~\cite{Hurd,Matl}.  
Typically, $R_s$ starts out small at low $T$, increases rapidly
as a power-law of $T$ to attain a broad peak at $\sim$0.8 $T_C$.
As $T$ crosses $T_C$, $R_s$ decreases gradually into the paramagnetic state.  
Often, but not always, $R_s$ changes sign just below $T_C$, as it does
here (Fig. \ref{RsRH}a).

The isolation of $\sigma^{in}_{xy}(T)$, which is opposite in sign to the KL term
and strictly monotonic in $T$, clarifies significantly the $T$ profile of $R_s(T)$.  
By Eqs. \ref{Rs} and \ref{sAM}, we have
\be
R_s(T) = \rho(T)^2\left[ S_H + \frac{\sigma^{in}_{xy}}{M}\right].
\label{RsT}
\ee
At low $T$, $\sigma^{in}_{xy}$ is negligible.  With the assumption that $S_H$
is a constant, $R_s$ initially increases as $\rho^2$.  As $\sigma^{in}_{xy}/M$ 
grows to dominate $S_H$ in the interval 80-120 K, the quantity
within $[\cdots]$ steadily decreases, changing sign near 125 K.  The
steep increase in the prefactor $\rho^2$ causes $R_s$ to go through a broad peak
before plunging to large negative values.  Hence the simple dependence
$\sigma^{in}_{xy}\sim(\Delta\rho)^2$ directly accounts for the profile
of $R_s(T)$.

The angular dependence of the AHC is shown in Fig.
\ref{sxyangle} (at 5 K where inelastic excitations are negligible).  
The size of the jump $\Delta\sigma (\theta)$ is nominally independent
of $\theta$, consistent with the results in Fig. \ref{figsxy}b.  
Since the AHC is determined only by $\bf M$, it
follows that the direction and magnitude of $\bf M$ are nominally
independent of the field tilt angle $\theta$.  By contrast, the 
$H$-linear segments which represent the OHE
scale as $\cos\theta$, i.e. the component of $H$ normal to $\bf I$.  An interesting 
feature is seen in the curve at $\theta = 75 ^{\circ}$. The
rounding of the corner of the hysteretic loop is attributed to
reversible rotation of $M$ just below the coercive field \cite{JMMM-69-149}.


\section{Magnetoresistance}\label{secMR}
The MR displays a rich assortment of behaviors depending on the 
field geometry.  The MR curves measured with the current $\bf I$ in the
$ab$-plane are displayed in Figs. \ref{MRall}a and \ref{MRall}b
in the configurations $\bf H\parallel c$ and $\bf H \perp c$, respectively. 
In the latter, we align $\bf H \parallel I$ to minimize 
carrier deflections by the Lorentz force.
In Panel (a), the MR is always negative and nominally $H$-linear. The
relative decrease is largest just below $T_C$, but becomes much weaker
as $T\rightarrow$ 10 K.  Below 100 K, the jumps in $M$ at the coercive fields $H_c$ produce jumps in $\rho$ that become more prominent at low $T$ (the jumps are
discussed in Sec. \ref{secjump}).  By sharp contrast, the MR in 
Panel (b) shows quite the opposite trend as $T$
decreases below $T_C$.  Below 100 K, the MR is positive and 
quadratic in $H$.  Its magnitude increases
dramatically as $T\rightarrow$ 10 K.

\bfig
\incl[width=9.5cm] {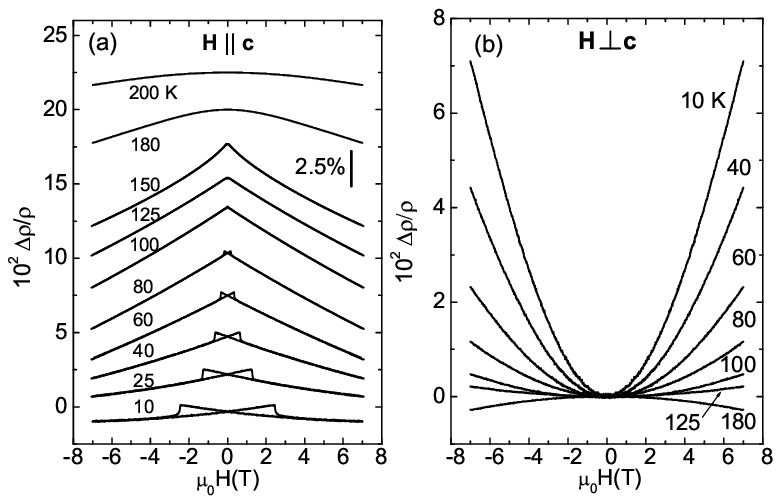}
\caption{\label{MRall} 
Magnetoresistance curves at selected temperatures in the 
field geometry $\bf H\parallel c$ (Panel a) and $\bf H\perp c$ (b).
In Panel (a), the MR is negative and nominally $H$-linear
indicating the dominance of the magnon-suppression mechanism.
The curves are displaced vertically by 2.5$\%$ for clarity.  
The bow-tie features correspond to jumps $\delta\rho$ at $H_c$
(Sec. \ref{secjump}).
In (b), the MR is positive with $H^2$ variation, reflecting the
AMR effect. The current $\bf I\perp c$ in both panels.
}
\efig

\begin{figure}
\incl[width=9.3cm]{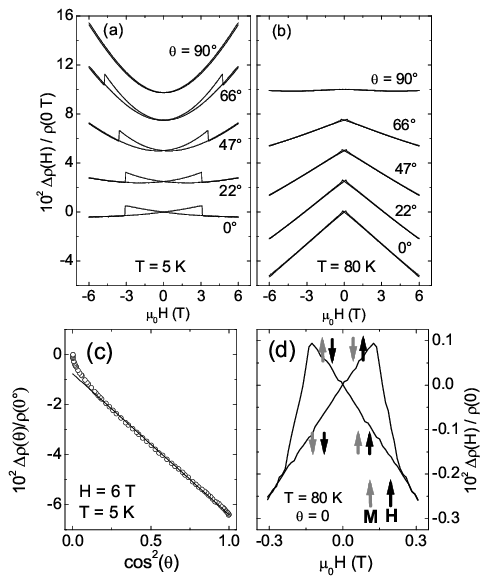} 
\caption{\label{MRangle} 
The MR curves measured at selected field tilt-angles
$\theta$ = 0, $\cdots$, 90$^\circ$ (with $\bf I\perp c$)
at 5 K (Panel a) and 80 K (b). Curves are displaced
by 2.5$\%$ for clarity.  As $\theta$ increases in Panel (a),
the AMR mechanism is increasingly dominant (positive MR and $H^2$).  
However, in (b), the magnon-suppression mechanism,
dominant at $\theta = 0$, progressively weakens as 
$\theta \uparrow 90^{\circ}$.
In Panel (c), the MR measured at 5 K in $H$ = 6 T is plotted
against $\cos^2\theta$.  The solid line is a fit to Eq. \ref{AMR}.
Panel (d) is an expanded view of the bow-tie 
hysteresis loop of $\rho$ caused by the jumps in $\bf M$, measured
with $\bf H\parallel c$ at $T$ = 80 K.  Black (grey) arrows represent
directions of $\bf H$ ($\bf M$) at selected segments of loop.
}
 \end{figure}

We have investigated the angular dependence of the MR to learn
more about the MR.  Figure \ref{MRangle}a shows the MR curves measured at 5 K 
for a series of tilt angles $\theta$ = $0, \cdots, 90^\circ$ 
($\theta$ is the angle between $\bf H$ and $\bf c$; see
inset in Fig. \ref{sxyangle}).  With increasing tilt,
the MR curve, initially negative, becomes strongly positive
with a $H^2$ dependence.  However, the same sequence
of measurements at 80 K (Fig. \ref{MRangle}b) 
shows a different trend.  The slope of the nominally linear curves
at $\theta = 0$ weakens substantially as $\theta\rightarrow 90^\circ$.
The overall patterns in Figs. \ref{MRall}a and \ref{MRall}b,
and Figs. \ref{MRangle}a and \ref{MRangle}b suggest the existence
of 2 competing MR contributions, one that is positive with an $H^2$ variation
and the other that is $H$-linear and negative.

\subsection{Anisotropic Magnetoresistance}\label{secAMR}
At low $T$, a dominant contribution to the MR is the 
anisotropic magnetoresistance (AMR) effect, in which the resistivities
$\rho_{\parallel}$ and $\rho_{\perp}$ (measured 
with $\bf I \parallel M$ and $\bf I\perp M$, respectively) differ measurably.
In most experiments, the difference
$\rho_{\Delta} \equiv  (\rho_{\parallel} - \rho_{\perp})$ is found to be positive.
The AMR has been explained~\cite{Smit,Fert,McGuire} by an anisotropy in the scattering of carriers in the $s$ band to a $d$ state and back to the $s$ band without spin flip,
$$|4s,{\bf k}\downarrow\rangle \rightarrow |3d\downarrow\rangle \rightarrow |4s,{\bf k'}\downarrow\rangle.$$
The spin-orbit term $\lambda {\bf L\cdot S}$ leads to mixing of the
$d$ spin states $|d\uparrow\rangle$ and $|d\downarrow\rangle$, while the direction
of $\bf M$ imparts a vector direction that, in effect, enhances the scattering amplitude
for electrons moving with velocity $\bf v\parallel M$ over those with $\bf v\perp M$.  This anisotropic selectivity, intrinsically 
tied to $s-d$ transitions that do not flip spin, tends to be 
suppressed when inelastic scattering processes that flip the spin
are important.  Hence, the AMR mechanism weakens rapidly in the 
presence of magnon scattering at elevated $T$.

\begin{figure}
\incl[width=7cm]{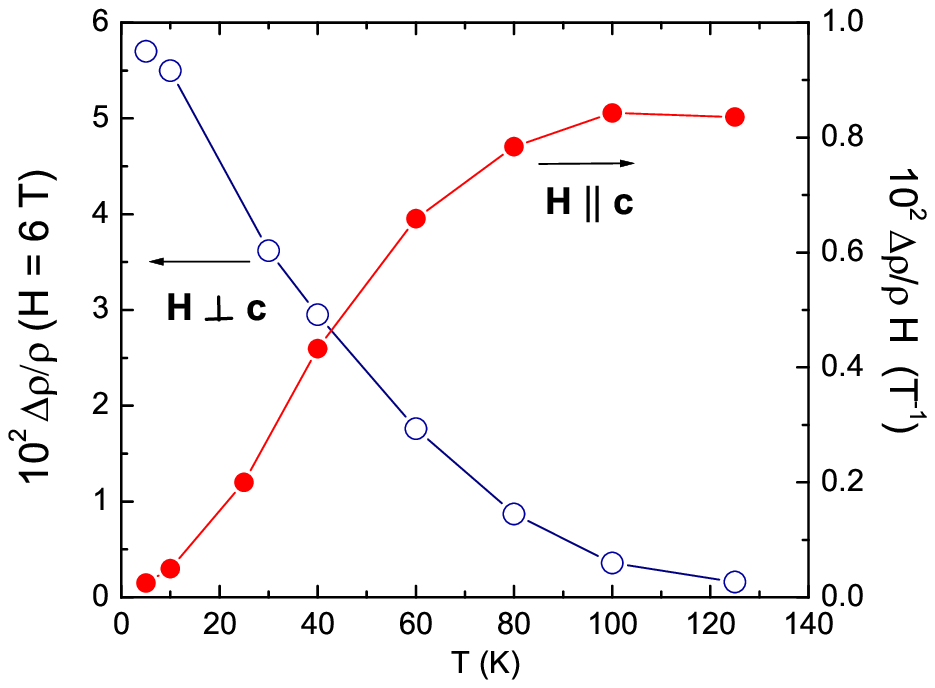} 
\caption{\label{MRvsT}  (Color online)
The $T$ dependence of the MR signals measured with
$\bf H\perp c$ (open circles) and $\bf H\parallel c$ (closed).
For $\bf H\perp c$, the MR signal is the fractional change $\Delta\rho(T,H)/\rho(T,0)$ measured at 6 T.  For $\bf H\parallel c$,
we have plotted the initial slope of the fractional change 
$\Delta\rho(T,H)/\rho(T,0)H$ ($H\rightarrow 0$).
}
\end{figure}

In the field-tilt experiments, we alter 
the respective fractions of carriers 
with $\bf v\parallel M$ and $\bf v\perp M$
by changing the tilt-angle $\alpha$ of $\bf M$ relative to $\bf c$ 
(we keep $\bf I\perp c$; see inset in Fig \ref{sxyangle}))~\cite{Gil}.
To leading order in $\alpha$, the AMR is expressed as
\be
\rho(\alpha) = \rho_{\perp} + \rho_{\Delta}\sin^{2}(\alpha).
\label{rhoalpha}
\ee

It is more convenient to express the MR in terms 
of the angle $\theta$ between $\bf H$ and $\bf c$.
Using $\chi_{ab}$ to eliminate $\alpha$, we have
\begin{equation} 
\rho(\theta) = \rho_{\perp} +
\rho_{\Delta}\biggl[ \frac{\chi_{\textrm{ab}}H}{M_{\textrm{s}}}\biggr]^{2} -
\rho_{\Delta}\biggl[\frac{\chi_{\textrm{ab}}H}{M_{\textrm{s}}}\biggr]^{2}\cos^{2}(\theta),
\label{AMR}
\end{equation}
where $M_{s}$ is the saturated magnetization.  The fit of Eq. \ref{AMR}
to the MR data taken at 5 K with $H$ = 6 T is quite good 
(Fig. \ref {MRangle} (c)).  The fit yields a large
$\rho_{\Delta} = +260 \;\mu\Omega$cm that is more than 5$\times$
$\rho$($H$ = 0 T) (the positive sign of $\rho_{\Delta}$ is similar
to that in most ferromagnets).  The AMR effect 
leads to positive MR with an $H^2$ dependence up to at least 14 T.
Consequently, the AMR is dominant in the MR curves with $\bf H$ at a sizeable tilt angle $\theta$ (Fig. \ref{MRall}b and \ref{MRangle}a).

The $T$ dependence of the MR data is displayed in Fig. \ref{MRvsT}.
In the geometry with $\bf H\perp c$ (open circles), the MR
signal, reported as the fractional increase $\Delta\rho(T,H)/\rho(T,0)$
with $H$ fixed at 6 T, 
decreases sharply from 5.8$\%$ at 5 K to 0.1$\%$ at 120 K.  This
is as expected if the AMR mechanism dominates the MR in this geometry.

By contrast, with $\bf H\parallel c$,
$\bf M$ remains pinned to $\parallel\bf c$, so 
the AMR is very weak.  The MR is then dominated by
the magnon suppression mechanism
discussed in the next section (Sec. \ref{secmagnons}).  
The (negative) MR signal increases with
$T$, consistent with magnon-suppression 
(closed circles in Fig. \ref{MRvsT}). 

\begin{figure}
\incl[width=7cm]{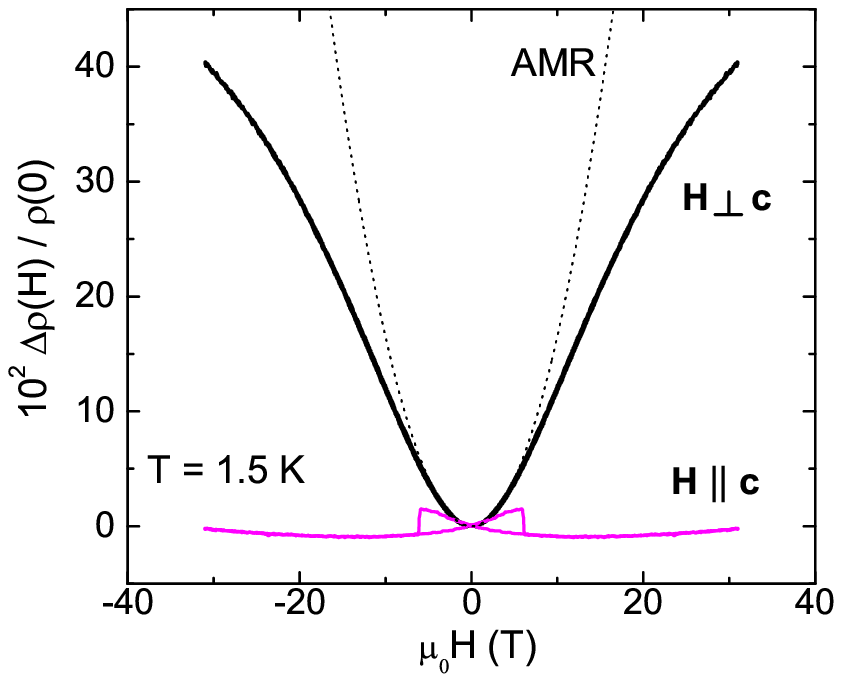}
\caption{\label{highH} (color online) 
Magnteoresistance measured to 31 T with $\bf H\perp c$ (bold curve)
and $\bf H\parallel c$ (faint curve with bow-tie feature).  
The dashed curve is the weak-field AMR expression Eq. \ref{AMR}.
}
\end{figure}

We have also made limited MR measurements in high fields,
up to 31 T (Fig. \ref{highH}).  The high field MR with $\bf H\perp c$
taken at 1.5 K (bold curve) shows 
significant deviation from the $H^2$ trend of 
Eq. \ref{AMR} (dashed curve) above 10 T, confirming that the
in-plane susceptibility $\chi_{ab}$ gets smaller as the tilt
angle $\alpha$ of $\bf M$ gets large, as expected.  From the
40 $\%$ change in $\rho$ at 31 T, we calculate that 
$\alpha \sim 15^{\circ}$.  In the geometry $\bf H\parallel c$
(faint curve), the MR is negative and shows a prominent bow-tie,
but is quite small overall.  Above 15 T, we detect an $H^2$ 
upturn associated with orbit bending by the Lorentz force.  This curve
shows the field scale needed to detect the classical orbital 
contribution to the MR in perpendicular field.

\subsection{Field suppression of magnons}\label{secmagnons}
The second important contribution to the MR in ferromagnets is 
the field suppression of the magnon population. 
In applied field, the Zeeman energy stiffens the 
restoring force against thermally-induced 
fluctuations of the moments away from 
equilibrium, thereby raising the energies of magnon branches. The
consequent decrease in magnon population reduces scattering of the carriers. 
The effect depends only on the component of $\bf H$ $\parallel {\bf M}$.
Hence, for $\bf H\parallel M$, one observes a negative MR with $H$-linear variation (reflecting the $H$-linear increase in Zeeman energy)~\cite{Raquet,Gil}.  
The negative MR curves observed in the geometry $\bf H\parallel c$
are consistent with this magnon-suppression mechanism. 
In Fig. \ref{MRall}a, the fractional decrease $\Delta\rho/\rho$ at say 6 T
rises rapidly from 5 to 150 K.  We quantify the negative, $H$-linear
MR by the initial slope of the fractional change in $\rho$, viz. $\Delta\rho(T,H)/\rho(T,0)H$.  Its $T$ dependence, revealing a sharp increase
in the interval 5--100 K, is consistent with the increased dominance of
magnon scattering at elevated $T$ (solid
circles in Fig. \ref{MRvsT}).  As mentioned, this trend is opposite to that
of the AMR effect (open circles).

The reversal in dominance of the 2 mechanisms is also evident in
the field-tilt experiment (Fig. \ref{MRangle}).  
At low $T$ (Panel a), increasing $\theta$
converts the weak, negative MR (weak magnon suppression) to a large 
positive $H^2$ MR (dominant AMR).  However, at high $T$ (80 K, Panel b), 
the opposite is observed.  The magnon-suppression is dominant 
at $\theta =0$, but nearly unresolved at 90$^{\circ}$.

\subsection{Resistance jumps at coercive field}\label{secjump}
At low $T$, as $\bf H$ is gradually ramped from, say, $+\bf z$ to $-\bf z$, $\bf M$ remains pinned along $+\bf z$ until $H = -H_c$; 
the magnetization is trapped in a metastable 
configuration by the large anisotropy.  The abrupt 
reversal of $\bf M$ at $-H_c$ causes a jump in the resistivity $\delta\rho$,
which is visible in all MR curves except those at $\theta = 90^{\circ}$.
The hysteresis in the MR curve has a bow-tie feature
as shown in Fig. \ref{MRangle}d.  The arrows indicate the
directions of $\bf H$ and $\bf M$ on the segments just before 
and after a jump.

Both the AMR and magnon-suppression mechanisms are expected to contribute
to the jump $\delta\rho$.  At elevated $T$, the latter is
dominant.  Interestingly, in the metastable state, with $\bf -M\parallel H$,
the Zeeman energy serves to \emph{soften} the restoring force
against deviations $\delta{\bf M}$, so that the magnon population
is enhanced by field.  Thus, as $H$ increases in the negative direction,
$\rho$ increases until reversal of $\bf M$ occurs at $H = -H_c$.  
Then, the magnons adjust to the equilibrium population causing
$\rho$ to jump downwards.  This agrees with the pattern seen in
the curves taken with $\bf H\parallel c$ at high $T$.

At very low $T$, however, the exponential decrease in the magnon
population renders the above mechanism ineffectual.  Yet, we observe
the jump magnitude $\delta\rho$ to increase to large values at 0.3 K
(see curve at 1.5 K in Fig. \ref{highH}).  In the limit $T\rightarrow 0$, we reason that the large $\delta\rho$ is primarily caused by the AMR effect, even though AMR is insignificant when $|H|>H_c$ (for $\bf H\parallel c$).
In the metastable state (e.g. with $\bf H\parallel -z$, $\bf M\parallel +z$), the demagnetization field
exerts a strong force on the pinned magnetization, particularly near
the edge of the crystal.  This causes the magnetization vector $\bf M(r)$
to splay, producing a weak gradient in tilt angle $\alpha({\bf r})$ 
($\alpha({\bf r})$ varies from 0 in the center to a value $\alpha_0$ at the edge).  Although the average $\langle \alpha\rangle = 0$ over the sample volume, the mean of the square $\langle\alpha^2\rangle$ is finite. 
By Eq. \ref{rhoalpha}, the gradient leads to an AMR.  From the measured $\delta\rho/\rho\sim 1.5\%$, and the magnitude of $\rho_{\Delta}$ obtained above, we estimate that $\alpha_0\sim 0.1^{\circ}$.  Thus, a very slight gradient in $\bf M(r)$
is sufficient to account for the jump magnitude at 0.3 K.  
After the jump, $\bf M$ is in equilibrium everywhere. With the
vanishing of the gradient, the AMR is insignificant, as discussed
above for this geometry.

\subsection{Summary of magnetoresistance}
The detailed results on the MR in Fe$_{\frac14}$TaS$_2$ 
at various tilt angles and over 
a broad interval of $T$ uncover a rich pattern of behavior,
which we show is consistent with the existence of 
2 spin-related mechanisms, AMR and magnon-suppression.
The classical Lorentz-force mechanism is insignificant until $H$ exceeds $\sim$
15 T (at low $T$).  Although the arguments are largely qualitative, the broad
range of measurements provides a fairly objective test that we find
highly persuasive.
In the geometry $\bf H\perp c$, the AMR is dominant with
magnon-suppression unobserved, so the MR is positive, increasing
as $H^2$.  The suppression of AMR with increasing inelastic excitaions
accounts for the steep fall of the MR at high $T$ in this geometry.
However, with $\bf H\parallel c$,
the magnon-suppression mechanism is dominant while AMR is inoperative.
Accordingly, the MR is negative and nominally $H$-linear.  The increase in the MR signal with $T$ is consistent with the magnon suppression
mechanism. Finally, the jump magnitude of the resistance at $H_c$ is
explained by the combination of these 2 mechanisms.

In relation to the AHE and the inelastic term $\sigma^{in}_{xy}$ inferred in Sec. \ref{secAHE}, the MR results (with $\bf H\parallel c$) 
provide clear evidence that carrier scattering by 
magnons is the dominant factor determining the strong
$T$ dependence of the resistivity in the interval 20--100 K.
This implies that the increase of $\sigma^{in}_{xy}$ with $T$,
which scales as $(\Delta\rho)^2$, originates from carrier scattering 
by the magnons at low $T$ and other spin excitations nearer to $T_C$.

\section {Discussion}\label{discuss}
In Fe$_{\frac14}$TaS$_{2}$, 
the jump in $\bf M$ provides a way to accurately determine the
anomalous Hall conductivity.  From analysis of 
the $T$ dependence of the AHC, we find that it is $T$
independent from 5 to 50 K, despite an increasing $\rho$.
This behavior is consistent with the Berry-phase/KL theory
which predicts that the AHC is dissipationless (independent
of the carrier transport lifetime $\tau$).  Any residual
$T$ dependence comes from $M(T)$ via Eq. \ref{sigmaA}.  In a few
ferromagnets, La$_{1-x}$Sr$_x$CoO$_3$~\cite{Onose} and MnSi~\cite{mlee07}, this is found to
be the case over an extended interval of $T$ up to the $T_C$. 
However, in the majority of ferromagnets, the AHE coefficient
$R_s(T)$ displays a strong $T$ dependence (and often a sign reversal),
which indicates a more complicated picture at elevated $T$.  
Despite the accumulating evidence in support of
the validity of the Berry-phase/KL theory at low $T$, 
the role of inelastic excitations is poorly understood.

Here we find that a negative, inelastic term
$\sigma^{in}_{xy}$ becomes resolved above 50 K, and increases rapidly
in magnitude as $T\rightarrow T_C^{-}$.  We show that the non-monotonic
complicated $T$ profile of $R_s$ is accounted for as a 
sum of this term and a positive KL term $\sigma^{KL}_{xy}$
(Eq. \ref{RsT}).  As mentioned, the isolation of $\sigma^{in}_{xy}$
rests on the assumption that the KL term is strictly
given by $S_HM(T)$ with $S_H$ a constant.  

As shown in Fig. \ref{sAvsT}b, the inelastic AHC scales as $(\Delta\rho)^2$.  
Expressed in terms of $\tau$, we have $\sigma^{in}_{xy}\sim 1/\tau^2$.
This rules out, as the origin of the
inelastic term, skew scattering~\cite{skew,Hurd} which scales as
$\sigma^{skew}_{xy}\sim \tau$.  

We sketch our ideas on interpreting the inelastic AHC term $\sigma^{in}_{xy}$.
Its strong $T$ dependence suggests that it is more insightful to
view this term as a transverse current arising
from scattering off spin excitations, whose density $n_s(T)$
rises rapidly with $T$.  (At low $T$, these are magnons.  
However, as $T\rightarrow T_C^{-}$, the large
fluctuations in the order parameter render the spin-wave picture invalid,
and singular or large-amplitude fluctuations involving spin textures dominate.
We assume $n_s(T)$ includes both spin waves and these textures.)
From the MR results in Sec. \ref{secMR},
the $T$ dependence of $\rho$ is dominated by scattering 
from spin excitations, i.e. $\Delta\rho\sim n_s(T)$.  Our finding
then implies $\sigma^{in}_{xy}\sim n_s(T)^2$.  

If each spin excitation
generates independently a contribution to the Hall current, 
we would have observed a linear dependence on $n_s$.  
The higher power $n_s^2$ implies that the Hall current
selectively responds to specific correlations between spin excitations.  
In the Hall studies on manganite~\cite{Matl} and pyrochlore~\cite{Tokura}, 
it was argued that a Berry-phase Hall current is produced by correlations 
between local moments.  Hopping of an electron between 3 local moments
(core spins) that subtend a finite solid
angle $\Omega$ (i.e. finite chirality ${\bf S}_1\cdot{\bf S}_2\times{\bf S}_3$) forces
the electron spin $\bf s$ to describe the same solid angle $\Omega$ for strong
Hund coupling.  By Berry's argument, the electron acquires 
a geometric phase $\chi_B\sim\frac12\Omega$, 
which translates into a large Hall current (this is distinct 
from the KL AHE).  In effect, the Hall current is a chirality detector.  
Likewise, we may expect
that scattering of Bloch-state electrons from fluctuating spins leads to a large Hall
current that selectively detects regions with large average chirality
$\langle{\bf S}_i\cdot{\bf S}_j\times{\bf S}_k\rangle$. As a consequence, it 
measures chiral correlations in the strongly fluctuating spins.  This mechanism is
analyzed in the calculations in Refs. \cite{Ye,Tatara}.

The results in Fig. \ref{sAvsT}b suggest that the AHE in the 
inelastic regime in Fe$_{1/4}$TaS$_2$ involves the Berry phase arising from 
scattering from spin fluctuations with finite chirality.  The possible ubiquity of the
simple scaling relationship $\sigma^{in}_{xy}\sim (\Delta\rho)^2$ in ferromagnets 
will be explored in future studies.

\vspace{3mm}
\centerline{*   *   *}
\vspace{3mm}

We have benefitted from discussions with N. Nagaosa,
S. Maekawa and S. Onoda.
Research at Princeton University was supported by the U.S.
National Science Foundation (NSF) under grant DMR 0213706.  
The high-field measurements were performed at the National 
High Magnetic Field Lab., Tallahassee, a
national facility supported by NSF and the State of Florida.

\begin{figure}
\incl[width=7cm]{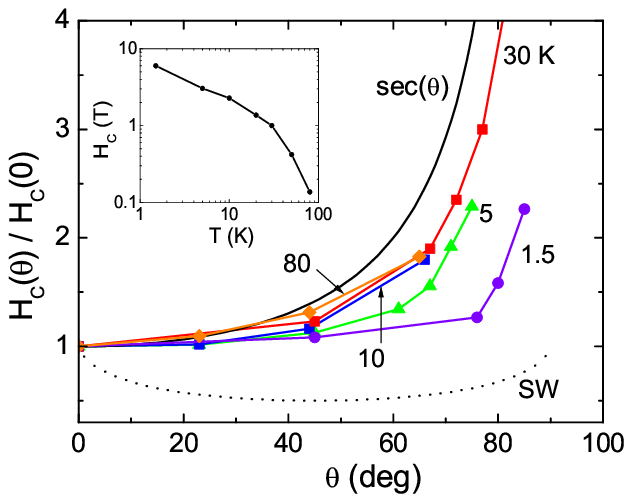}
 \caption{\label {Hcangle} (Color online) 
(Main Panel) 
The angular dependence of the measured 
coercive field $H_c$ at selected $T$.  
The predicted curve for $H_c$ based on the domain 
wall-mediated reversal of $M$ is labeled $\sec{\theta)}$.  The dashed
curve SW is the coherent-rotation expression.
The inset shows the $T$ dependence of $H_c$ plotted in 
log-log scale for $\theta = 0$. 
}
\end{figure}
\appendix
\vspace{5mm}
\centerline{\bf Appendix: Angular dependence of coercive field}
\vspace{5mm}
The small value of the ratio $\frac{H_{c}}{H_{A}}\sim$ 0.06--0.1 
suggests that the jumps in $\bf M$ at $H_c$ are triggered by
the rapid motion of domain walls.  The initial 
$M$-$H$ curve (taken after zero-field cooling) which shows that $M\simeq$ 0
until $H$ nears $H_{c}$ also suggests that depinning of pinned domain walls causes the jumps~\cite{Domains}. Models of domain-wall mediated reversals
predict $H_c(\theta)/H_c(0) \sim \sec(\theta)$ (the component
of $\bf H||M$ drives domain-wall motion). By contrast, in 
coherent rotation of $\bf M$ (as in small particles), the 
Stoner-Wohlfarth (SW) model predicts~\cite{SW,Givord}
$H_{c}(\theta) = H_{c}(0)[\sin^{2/3}(\theta) +
\cos^{2/3}(\theta)]^{-3/2}.$
Values of $H_c(\theta)/H_c(0)$ inferred from our tilt experiments
are plotted in Fig. \ref{Hcangle} and compared with $\sec(\theta)$
(solid curve) and the SW model (dashed curve).  Although the results above
10 K are consistent with the $\sec(\theta)$ curve, there is strong 
deviation at low $T$.  It appears that a large population of
thermal excitations is needed to bring $H_c(\theta)$ into 
agreement with the domain-wall model.  The trend suggests that, in the limit $T\rightarrow 0$, $H_c$ becomes \emph{independent} of $\theta$ 
over the interval $0<\theta<70^{\circ}$. 
Tunneling processes may become dominant
at low $T$.  The $T$ dependence of $H_c$ (inset) shows that $H_c$ approaches
saturation although at a gradual rate.  This is an interesting regime of coercivity dynamics that is little investigated.

\end{document}